\documentclass[prc,twocolumn,aps,tightenlines,showpacs,floatfix,nolongbibliography,nofootinbib,lengthcheck]{revtex4-2}
\usepackage{bm}
\usepackage{color}
\usepackage{graphicx,epsfig}
\usepackage{amsfonts,amssymb,amsmath}
\usepackage{longtable}
\usepackage{xcolor}
\usepackage{color}

\usepackage{dcolumn}     

\newcommand{\bfr}{{\bf r}}

\newcommand{\bfk}{{\bf k}}
\newcommand{\bfq}{{\bf q}}

\newcommand{\bfx}{{\bf x}}

\newcommand{\bfzh}{\hat{\bf z}}
\newcommand{\mel}[3]{\langle #1 | #2 | #3 \rangle}
\newcommand{\ket}[1]{| #1 \rangle}

\def\beq{\begin{equation}}
\def\eeq{\end{equation}}
\def\beqy{\begin{eqnarray}}
\def\eeqy{\end{eqnarray}}

\begin{document}

{
\title{Partial muon capture rates in $A=3$ and $A=6$ nuclei with chiral effective field theory}
\author{G.~B. \ King$^{\rm a}$, S.\ Pastore$^{\rm a,b}$, M.\ Piarulli$^{\rm a,b}$, 
 R.\ Schiavilla$^{\rm c,d}$}
\affiliation{
$^{\rm a}$\mbox{Department of Physics, Washington University in Saint Louis, Saint Louis, MO 63130, USA}
$^{\rm b}$\mbox{McDonnell Center for the Space Sciences at Washington University in St. Louis, MO 63130, USA}
$^{\rm c}$\mbox{Department of Physics, Old Dominion University, Norfolk, VA 23529, USA}
$^{\rm d}$\mbox{Theory Center, Jefferson Lab, Newport News, VA 23606, USA}
}
\date{\today}
\begin{abstract}
Searches for neutrinoless-double beta decay rates are crucial in addressing 
questions within fundamental symmetries and neutrino physics. The rates of these decays depend 
not only on unknown parameters associated with neutrinos, but also on nuclear properties. In order to
reliably extract information about the neutrino, one needs an accurate treatment of the complex many-body
dynamics of the nucleus.
Neutrinoless-double beta decays take place at momentum transfers on the order of 100 MeV/$c$ and require both nuclear electroweak vector 
and axial current matrix elements. 
Muon capture, a process in the same momentum transfer regime, has readily available 
experimental data to validate these currents. 
In this work, we present
results of {\it ab initio} calculations of partial muon capture rates for $^3$He and $^6$Li nuclei using variational and
Green's Function Monte Carlo computational methods. We estimate
the impact of the three-nucleon interactions, the cutoffs used to regularize
two-nucleon ($2N$) interactions, and the energy range of $2N$ scattering data used to fit these interactions.
\end{abstract}


\maketitle
}

{\it Introduction, conclusions and outlook.}  Nuclei play a crucial role in high-precision tests of the Standard Model and searches 
for physics beyond the Standard Model. These investigations, including neutrinoless 
double beta decay ($0\nu\beta\beta$) searches~\cite{Engel:2016xgb} and high-precision beta decay experiments~\cite{Gonzalez-Alonso:2018omy,Cirigliano:2019bd,Glick-Magid:2021uwb}, 
require a thorough understanding of standard nuclear effects in order to separate them from new physics signals.
In particular, $0\nu\beta\beta$ decay experiments aim to establish the origin and nature
of neutrino masses and test leptogenesis scenarios leading to the observed matter-antimatter
asymmetry in the universe~\cite{Engel:2016xgb}. Rates of these decays depend not only on 
unknown neutrino parameters but also on nuclear matrix elements.
The latter can be provided only from theoretical calculations. Thus, a prerequisite to this experimental 
program is an accurate treatment of the complex many-body dynamics of the nucleus and 
its interactions with neutrinos. If one assumes that $0\nu\beta\beta$ decay results come from the exchange of a 
light Majorana neutrino between two nucleons, then the momentum carried by the neutrino is on the order
of 100 MeV/$c$~\cite{Engel:2016xgb,Pastore:2018ndb}. Muon captures on nuclei---processes where a muon 
captures on a proton in the nucleus releasing a neutron and a neutrino---involve
momentum transfers on the order of the muon mass. The scope of this work is to validate our nuclear model 
in this kinematic regime by calculating muon capture rates in $A\,$=$\,3$ and $A\,$=$\,6$ nuclei for comparison with
available experimental data. 

Muon capture reactions have been treated extensively from both the theoretical and experimental points of view
~\cite{Mukhopadhyay:1977,Measday:2001,Kammel:2010,Gorringe:2015} and rates have been obtained in light systems with
several methods~\cite{Marcucci:2001qs,Marcucci:2011hh,Marcucci:2011jm,Congleton:1993epm,Congleton:1996epm,Gazit:2008vm,Golak:2014,Adam:2012hb,
Kolbe:1994crpa,Marketin:2009qrpa,Samana:2011rpa,Giannaka:2015qrpa}. 
Here, we present calculations of partial muon capture rates 
using quantum Monte Carlo methods (QMC)~\cite{Carlson:2014vla}---both variational (VMC) and Green's function
Monte Carlo (GFMC) methods---to solve the nuclear many-body problem. QMC methods allow one to fully retain the complexity of many-body physics 
and have been successfully applied to study many nuclear electroweak properties over a wide range of energy and momentum transfer, including 
total muon capture rates in $^3$H and $^4$He~\cite{Lovato:2019mc}, low-energy electroweak transitions~\cite{Pastore:2009,Pastore:2011,Pastore:2012em,Pastore:2014em,Pastore:2018gt}, 
nuclear responses induced by electrons and neutrinos~\cite{Lovato:2018ncqe,Lovato:2020ccqe,Pastore:2019urn},
neutrinoless double beta decay matrix elements~\cite{Pastore:2018ndb,Cirigliano:2018,Cirigliano:2019,Wang:2019}, and 
matrix elements for dark matter scattering~\cite{Andreoli:2018etf}.

The Norfolk two-nucleon ($2N$) and three-nucleon ($3N$) (NV2+3) local chiral interactions~\cite{Piarulli:2014bda,Piarulli:2016vel,Baroni:2016xll,Baroni:2018fdn} 
have been used in combination with QMC methods to study static properties of light 
nuclei~\cite{Piarulli:2016vel,Piarulli:2017dwd,Lynn:2019rdt,10.3389/fphy.2019.00245,Gandolfi:2020pbj}, and 
in auxiliary-field diffusion Monte Carlo~\cite{Schmidt:1999lik}, Brueckner-Bethe-Goldstone~\cite{bbg1,bbg2} and 
Fermi hypernetted chain/single-operator chain~\cite{FR75,PW79} approaches to investigate the equation 
of state of neutron matter~\cite{Piarulli:2019pfq,Bombaci:2018ksa}. Reference~\cite{King:2020wmp}, 
a study which included the current authors, reports on Gamow-Teller (GT) matrix elements calculated for $A\le10$ nuclei using 
the NV2+3 models and their consistent axial-vector currents at tree-level from Refs.~\cite{Baroni:2016xll,Baroni:2018fdn,Baroni:2015uza}.
The study validated the many-body interactions and currents in the limit of vanishing momentum transfer. 
In the present work, we use the same nuclear Hamiltonians and axial currents, along with chiral
vector currents retaining loop corrections developed in Refs.~\cite{Pastore:2008,Pastore:2009,Pastore:2011,Schiavilla:2018udt}, to test the model at 
moderate momentum transfers on the order of $100$ MeV/c and to assess the sensitivity of partial muon capture rates to the dynamical input. 

In the $A\,$=$\,3$  system, we obtain an average rate for all Norfolk models of $\Gamma(A=3;{\rm VMC}) = 1512 \pm 32~{\rm s}^{-1}$ 
at the VMC level that agrees with the experimental result of $1496.0 \pm 4.0 ~{\rm s}^{-1}$~\cite{Ackerbauer:1998} within error bar. In
the $A\,$=$\,6$ system, the VMC partial capture rate of $\Gamma(A=6;{\rm VMC})\,$=$\,1243 \pm 59~s^{-1}$ is significantly slower than the available experimental 
data point of $1600^{+330}_{-129} ~{\rm s}^{-1}$~\cite{Deutsch:1968}, but falls into the range of previous theoretical estimates~\cite{Fujii:1959,Lodder:1965,Kim:1970,Delorme:1970,Donnelly:1973,Mukhopadhyay:1973,Junker:1983}.
We analyzed uncertainties due to ({\it i}) the choice of cutoffs used to regularize the NV2 interactions, 
({\it ii}) the energy range of $2N$ scattering data used to fit the model low-energy constants (LECs), ({\it iii}) two different versions
of NV3 interactions, {\it i.e.}, the non-starred model fit to the $nd$ scattering length and the 
trinucleon binding energies, and the starred model fit to the triton GT matrix element and the trinucleon binding energies, 
and ({\it iv}) a 10\% variation in the nucleonic axial radius. In the $A\,$=$\, 3$ system, the largest source of uncertainty comes from the choice
of $3N$ interaction model, 
while for $A\,$=$\,6$ we find that the uncertainty due to the $3N$ interaction is slightly less
than, but on the order of, the cutoff and energy range uncertainties. On average, there is a change
in the rate by $\pm 0.6\%$ when the axial radius is varied in the interval $r_A = [0.5859, 0.7161]$ fm.

We improved upon our VMC estimate by performing GFMC propagations using 
models NV2+3-Ia and NV2+3-Ia*, or Ia and Ia* for short,
in both the $A\,$=$\,3$ and $A\,$=$\,6$ systems.
These models share the same $2N$ interaction, but differ in how the $3N$ interaction is fit with $(c_D,c_E)\,$=$\,(3.666,-1.638)$ for Ia~\cite{Piarulli:2017dwd} and 
$(c_D,c_E)\,$=$\,(-0.635,-0.090)$ for Ia*~\cite{Baroni:2018fdn}.  
Model Ia, constrained by strong interaction data
only, achieves 1.5\% agreement with the experimental datum for $A\,$=$\,3$
with a calculated rate of $1519\pm 3~{\rm s}^{-1}$.  Its counterpart, model Ia*, 
constrained to both strong and electroweak data, 
underpredicts the experimental rate by a few percent.  For $A\,$=$\,6$, we find that 
the model Ia* propagation significantly decreases the rate due to the monotonic growth of the $^6{\rm He}$ ground state rms radius at early 
imaginary times.  By contrast, model Ia has a stable radius throughout the GFMC
propagation and the rate decreases by less than 1\%; nevertheless, it still 
underpredicts the experimental datum.

Given the large error bars on the $^6$Li datum and the wide range of values from past theoretical 
calculations, we advocate for renewed experimental and theoretical attention to this partial capture rate.
While in this letter we focus on $^3$He and $^6$Li to demonstrate the impact 
of this sort of study, there are other muon capture rates with available experimental data which the combination of QMC methods and NV2+3 chiral Hamiltonians
could be made to address with future development; examples are $^{10}{\rm B}$, $^{11}{\rm B}$, $^{12}{\rm C}$, $^{16}{\rm O}$, and $^{40}{\rm Ca}$~\cite{Measday:2001}.
Calculations of these rates, particularly for the heavier nuclei,
would be valuable in further validating the present {\it ab initio} approach in the kinematic regime relevant to neutrinoless double beta decay.

{\it Partial muon capture rate.}  The muon capture processes $^3$He$(\mu^-,\nu_\mu)^3$H and
$^6$Li$(\mu^-,\nu_\mu)^6$He 
are induced by the weak-interaction Hamiltonian $H_W$~\cite{Walecka:1975,Walecka:1995}
\begin{equation}
\mel{\bfk_{\nu},h_{\nu}}{H_{W}}{\bfk_{\mu},s_{\mu}} = \frac{G_V}{\sqrt{2}}\int d^3x e^{-i\bfk_{\nu} \cdot \bfx} \tilde{l}_{\sigma}(\bfx)j^{\sigma}(\bfx)\, ,
\label{eq:weak}
\end{equation}
where $G_V\,$=$\, G_F\cos\theta_C\,$=$\,1.1363 \times 10^{-5}\, {\rm GeV}^{-2}$ is the Fermi coupling constant extracted from analyses of 
superallowed $\beta$-decays~\cite{Hardy:2015},
$j^{\sigma}$ and $\tilde{l}_{\sigma}$ are the hadronic and leptonic four-current density operators~\cite{Marcucci:2001qs},
$s_{\mu}$ is the muon spin, $h_{\nu}$ is the neutrino helicity, and $\bfk_{\mu}$ and $\bfk_{\nu}$ are the muon and neutrino momenta, 
respectively. The value of $G_V$ adopted here is from a more recent analysis and is $\sim 1.1\%$ smaller
than that used in previous calculations based on the hyperspherical harmonics method with chiral currents from Ref.~\cite{Marcucci:2011jm}.

For a transition from an initial nuclear state $\ket{i, J_i\, M_i}$---where $J_{i/f}$ and $M_{i/f}$ denote the nuclear spin and 
its projection---to a final nuclear state $\ket{f,J_f\,M_f, -{\bf k}_\nu}$
recoiling with momentum $-{\bf k}_\nu$, the general expression for the capture rate ($\Gamma$), summed 
over the final states and averaged over the initial states, is given (in the limit of vanishing
$\bfk_{\mu}$) by~\cite{Walecka:1975,Walecka:1995,Marcucci:2001qs}
\begin{align}
& d\Gamma =  \frac{1}{2(2J_i +1)}\sum_{s_{\mu},M_i}\sum_{h_{\nu},M_f}2\pi\,\delta(\omega) \\
                   &\times |\mel{\bfk_{\nu},h_{\nu};f, J_f M_f,-{\bf k}_\nu }{H_{\rm{W}}}{s_{\mu}; i, J_i M_i} |^2\frac{d^3k_{\nu}}{(2\pi)^3} \ ,
                   \nonumber
\label{eq:diff.rate}
\end{align}
where the argument of the $\delta$-function is
\begin{equation}
\omega = E_{\nu} + \sqrt{E_{\nu}^2 + (m_f+E_f)^2} - \left( m_{\mu}+m_i+E_i \right) \ , 
\label{eq:econservation}
\end{equation}
and $E_i$ and $E_f$ are the initial and final state energies of the nucleus~\cite{Purcell:2015,Tilley:2002,Kelley:2017}---we
have neglected internal electronic energies, since they are of the order of
tens of eV's for the light atoms under consideration. 
We also used the following definitions
\begin{align}
m_i & = Z\,m_p + N\,m_n + (Z-1)\,m_e \, , \\ \nonumber
m_f & = (Z-1)(m_p + m_e) + (N+1)\,m_n \, , 
\end{align}
for an initial nucleus with charge number $Z$ and neutron number $N$, and we denoted with $m_p$, $m_n$, and $m_e$ 
the proton, neutron, and electron masses, respectively.

The final integrated rate can be conveniently written in terms of matrix
elements of the nuclear electroweak current components~\cite{Marcucci:2011hh},
\begin{widetext}
\begin{align}
\nonumber
\Gamma &= \frac{G_V^2}{2\pi}\frac{|\psi^{\rm{av}}_{1s}|^2}{(2J_i+1)}\frac{E_{\nu}^{*2}}{{\rm recoil}}\sum_{M_f,M_i} \left[ |\mel{J_f,M_f}{\rho(E^*_{\nu}\bfzh)}{J_i,M_i}|^2 + |\mel{J_f,M_f}{j_z(E^*_{\nu}\bfzh)}{J_i,M_i}|^2  \right.  \\ \nonumber
&+ 2{\rm Re}\left[ \mel{J_f,M_f}{\rho(E^*_{\nu}\bfzh)}{J_i,M_i}\mel{J_f,M_f}{j_z(E^*_{\nu}\bfzh)}{J_i,M_i}^* \right] + |\mel{J_f,M_f}{j_x(E^*_{\nu}\bfzh)}{J_i,M_i}|^2  \\[0.25cm]
& \left. + |\mel{J_f,M_f}{j_y(E^*_{\nu}\bfzh)}{J_i,M_i}|^2 - 2\,{\rm Im}\left[ \mel{J_f,M_f}{j_x(E^*_{\nu}\bfzh)}{J_i,M_i}\mel{J_f,M_f}{j_y(E^*_{\nu}\bfzh)}{J_i,M_i}^* \right] \right] \ ,
\label{eq:rate}
\end{align}
\end{widetext}
where we have chosen $\hat{\bfk}_{\nu}\,$=$\,-\bfzh$, and 
have introduced the outgoing neutrino energy~\cite{Marcucci:2011hh}
\begin{equation}
E^*_{\nu} = \frac{\left(m_i+E_i+m_{\mu}\right)^2 - \left(m_f + E_f\right)^2}{2\left(m_i+E_i+m_{\mu}\right)} \ , 
\label{eq:enustar}
\end{equation}
and recoil factor
\begin{equation}
\frac{1}{{\rm recoil}} =\left( 1 - \frac{E^*_{\nu}}{m_i+E_i+m_{\mu}}\right)\ .
\end{equation}
The factor $|\psi^{\rm av}_{1s}|^2$ is written as $ \mathcal{R} \,|\psi_{1s}(0)|^2$,
where $\psi_{1s}(0)$ is the 1s wave function, evaluated at the origin,
of a hydrogen-like atom, and ${\cal R}$ approximately accounts for the finite
size of the nuclear charge distribution~\cite{Marcucci:2011hh}, here calculated
with the NV2+3 Hamiltonians.

{\it Nuclear Hamiltonians and electroweak currents.} To calculate the nuclear matrix elements
required by Eq.~(\ref{eq:rate}) we employ VMC~\cite{Wiringa:1991kp} and GFMC~\cite{Carlson:1997qn} methods.
For a comprehensive review of these methods, see Refs.~\cite{Carlson:2014vla,Lynn:2019rdt}
and references therein. 
Details about the calculation of matrix elements using GFMC wave functions are found in Eqs.~(19)--(24) of Ref.~\cite{Pervin:2007}. 

The many-body Hamiltonian is composed of a (one-body) kinetic
energy term, and the Norfolk $2N$ and $3N$ local interactions that include N3LO and N2LO terms in the chiral expansion, respectively.
Details about the derivation of the interaction in chiral effective field theory can be found in
Ref.~\cite{Piarulli:2014bda,Piarulli:2016vel,Baroni:2016xll,Baroni:2018fdn}. Here, we briefly summarize the differences 
between the model classes employed in this work.
Models in class I (II) fit the $2N$ interaction to about 2700 (3700) data points
up to lab energy of 125 (200) MeV in the nucleon-nucleon scattering database
with a $\chi^2$/datum of about $\lesssim 1.1$ ($\lesssim 1.4$).  Within each class, models a and b differ
in the set of cutoffs adopted to regularize the short- and long-range components of the interaction, either $(R_S,R_L)\,$=$\,(0.8,1.2)$ fm
for model a or $(R_S,R_L)\,$=$\,(0.7,1.0)$ fm for model b~\cite{Piarulli:2014bda,Piarulli:2016vel}.
The different fitting procedures result in different values for the 26 unknown LECs governing 
the strength of short-range terms in the interactions.
Accompanying these $2N$ interactions is the leading chiral $3N$ interaction which
introduces two unknown LECs $c_D$ and $c_E$ (in standard notation) constrained to reproduce the
trinucleon binding energies and, concurrently, either the GT matrix element contributing to tritium $\beta$-decay~\cite{Baroni:2018fdn}
in the starred model or the $nd$-doublet scattering length in the non-starred one~\cite{Piarulli:2017dwd}.

Lastly, the vector- and axial-current operators entering the calculation were derived with time-ordered perturbation
theory by the JLab-Pisa group using the same $\chi$EFT formulation as the NV2+3 interactions.
Details about the electroweak currents used in this work can be found in
Refs.~\cite{Baroni:2016xll,Baroni:2018fdn,Baroni:2015uza,Pastore:2008,Pastore:2009,Pastore:2011,Schiavilla:2018udt}.

{\it Results.}  The results of the VMC calculation of the partial muon capture rate in $A\,$=$\,3$ and $A\,$=$\,6$ using the NV2+3 nuclear Hamiltonian are presented in Table~\ref{tab:rates}. 
Capture rates were determined using nuclear axial and vector current operators consistent with the NV2+3 model. The nuclear axial currents
\cite{Baroni:2016xll} contain only tree-level diagrams while the vector current operators account for loop corrections derived in 
Ref.~\cite{Pastore:2008,Pastore:2009,Pastore:2011,Schiavilla:2018udt}.

Calculations of the rate with the leading order one-body only [$\Gamma({\rm 1b})$] and one- plus two-body electroweak currents [$\Gamma({\rm 2b})$] were 
performed for ground-state to ground-state transitions. 
The partial capture rate on $^3$He
has been precisely measured~\cite{Ackerbauer:1998} and the one-body contribution alone cannot reproduce this measurement. With the 
two-body electroweak currents included, the VMC rates increase by about 9\% to 16\%. At this level, the agreement with the 
datum ranges from about 0.1\% to 4.6\%.  How the $3N$ interaction was fit
has the most significant impact on the rate, leading to differences on average of 54 ${\rm s}^{-1}$ whenever the
$3N$ interaction is changed.
Note that the LEC $c_D$ entering the $3N$ interaction governs the strength of the axial contact current at next-to-next-to-next-to-leading order 
in the chiral expansion~\cite{King:2020wmp}. Therefore, variations in the $3N$ interaction lead to variations in the current, as also 
observed in the study of Ref.~\cite{King:2020wmp} on beta decay matrix elements. The cutoff and 
energy range of the fit lead to changes of 16 ${\rm s}^{-1}$ and 22 ${\rm s}^{-1}$ on average, respectively, which is consistent with the 
findings of Refs.~\cite{Marcucci:2011jm,Gazit:2008vm}. 

In the $^6$Li capture,
the inclusion of two-body electroweak currents also increases the rate
with a greater enhancement in the non-starred models relative to their starred counterparts. 
Even with this increase, ranging approximately from $3\%~{\rm to}~7\%$, the rates predicted at the VMC level for the NV2+3 models are 
about 11--21\% slower than the available experimental datum ~\cite{Deutsch:1968}. Here, the difference due to
the $3N$ interaction is no
longer the dominant contribution to the uncertainty. We find that, on average, the cutoff and energy range of the fit both change the rate by 
$72~{\rm s}^{-1}$, while the $3N$ interaction changes the rate by $60~{\rm s}^{-1}$. 

We compute a VMC average for both rates under study and use the average changes due to
the chiral $2N$ interaction cutoffs, the energy range used to fit the interaction, and the $3N$ interaction to assign a total error bar. 
An additional source of uncertainty was considered by varying the nucleonic axial radius parameter by $\pm10\%$. We found that, on average, the 
difference in the rate was $\pm 0.6\%$ due to this variation. We combine the four uncertainties in quadrature to determine the overall uncertainty on the VMC averages, obtaining $\Gamma(A\!=\!3;{\rm VMC})\,$=$\,1512 \pm 32~{\rm s}^{-1}$ and
$\Gamma(A\!=\!6;{\rm VMC})\,$=$\,1243 \pm 59~{\rm s}^{-1}$.

\begin{table*}[tbh]
\begin{center}
\small
\begin{tabular}{ l  l  c  c c  c } \hline \\ [1pt]
Capture & Model & $\mathcal{R}$ & $\Gamma({\rm 1b})$ (s$^{-1}$) & $\Gamma({\rm 2b})$ (s$^{-1}$) & Expt. (s$^{-1}$)\\  [2pt] \hline
\\
$^{3}$He($\frac{1}{2}^+$;$\frac{1}{2}$) $\to$ $^{3}$H($\frac{1}{2}^+$;$\frac{1}{2}$) &Ia (Ib) & 0.995 (0.995) & 1350.3 $\pm$ 0.8 (1363.4 $\pm$ 0.2) &1564.4 $\pm$ 0.9 (1545.7 $\pm$ 0.3) & 1496.0 $\pm$ 4.0~\cite{Ackerbauer:1998} \\[2pt]
 &Ia* (Ib*) & 0.995 (0.995) &1357.4 $\pm$ 0.2 (1358.5 $\pm$ 0.2) & 1473.9 $\pm$ 0.3 (1483.6 $\pm$ 0.3) & \\[2pt]
&IIa (IIb) &0.995 (0.995) & 1369.7 $\pm$ 0.2 (1372.8 $\pm$ 0.2) &1533.8 $\pm$ 0.3 (1512.4 $\pm$ 0.3) &\\ [2pt]
&IIa* (IIb*) &0.995 (0.995) & 1364.5 $\pm$ 0.2 (1372.5 $\pm$ 0.2) & 1484.4 $\pm$ 0.3 (1497.0 $\pm$ 0.3) & \\ [2pt]
$^{6}$Li(1$^+$;0) $\to$ $^{6}$He(0$^+$;1) &Ia (Ib) &0.990 (0.990) &1196 $\pm$ 2 (1243 $\pm$ 2) &1282 $\pm$ 2 (1331 $\pm$ 2) & 1600$^{+330}_{-129}$~\cite{Deutsch:1968}\\[2pt]
&Ia* (Ib*) & 0.990 (0.990) & 1154 $\pm$ 3 (1188 $\pm$ 2) & 1177 $\pm$ 3 (1233 $\pm$ 2)  &\\[2pt]
&IIa (IIb) & 0.990 (0.990) &1227 $\pm$ 2 (1142 $\pm$ 2) &1294 $\pm$ 2 (1185 $\pm$ 2) &\\[2pt]
&IIa* (IIb*) &0.990 (0.990) & 1215 $\pm$ 2 (1151 $\pm$ 2) & 1257 $\pm$ 2 (1185 $\pm$ 2) & \\ [2pt]
\end{tabular}
\end{center}
\caption{VMC calculations of partial muon capture rates in $^3$He and $^6$Li obtained with chiral one-body only [$\Gamma({\rm 1b})$], and one- and two-body [$\Gamma({\rm 2b})$]
axial and vector currents corresponding to the eight NV2+3 models. The third column gives the factor $\mathcal{R}$
used to account for the finite nuclear charge distribution. The experimental result is given in the last column. All uncertainties on the theoretical predictions are Monte Carlo errors.}
\label{tab:rates}
\end{table*}

\begin{table*}[tbh]
\begin{center}
\small
\begin{tabular}{ l  l  c  c  c  } \hline \\ [1pt]
Capture & Model & $\Gamma$(VMC) (s$^{-1}$) &$\Gamma$(GFMC) (s$^{-1}$) &Expt. \\ [2pt] \hline
$^{3}$He($\frac{1}{2}^+$;$\frac{1}{2}$) $\to$ $^{3}$H($\frac{1}{2}^+$;$\frac{1}{2}$) &Ia &1564.4 $\pm$ 0.9 & 1519 $\pm$ 3 &1496.0 $\pm$ 4.0~\cite{Ackerbauer:1998}\\
 &Ia* &1473.9 $\pm$ 0.3 &1433 $\pm$ 2 & \\
$^{6}$Li(1$^+$;0) $\to$ $^{6}$He(0$^+$;1) &Ia &1282 $\pm$ 2 & 1277 $\pm$ 10 &1600 $^{+330}_{-129}$~\cite{Deutsch:1968}\\ 
 &Ia* &1177 $\pm$ 2 & 926 $\pm$ 8 &\\
\end{tabular}
\end{center}
\caption{VMC and GFMC calculations of partial muon capture rates in $^3$He and $^6$Li obtained with chiral one- and two-body  axial and vector currents 
with the NV2+3 models. The experimental result is given in the last column. All uncertainties on the theoretical predictions are Monte Carlo errors.}
\label{tab:gfmc.rates}
\end{table*}

\begin{figure}[tbh]
\begin{center}
\includegraphics[width=0.49\textwidth]{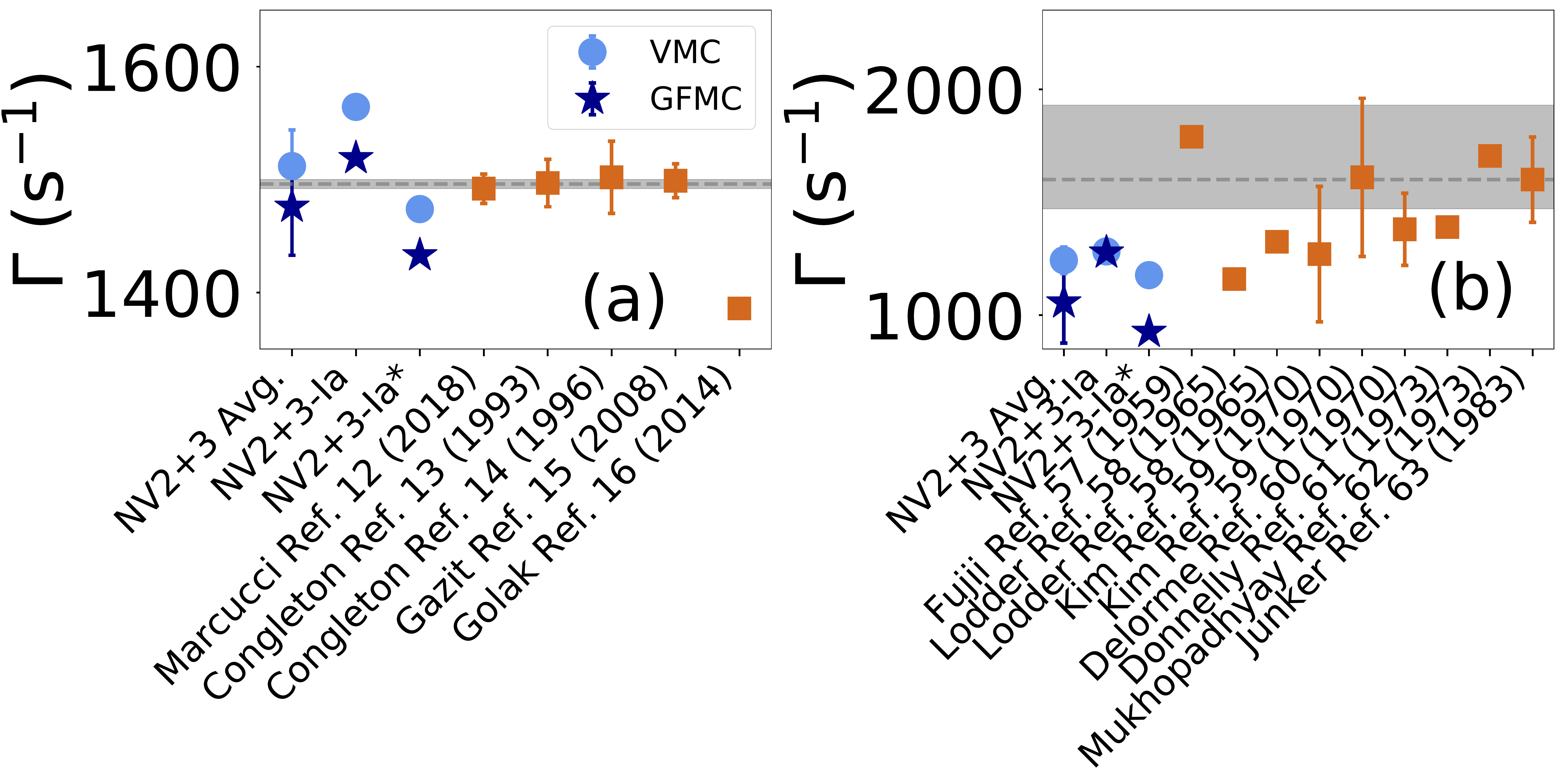}
\end{center}
\caption{The partial muon capture rate in (a) $^3$He and (b) $^6$Li from the the NV2+3-Ia and NV2+3-Ia* models in VMC (light blue circle) and GFMC (dark blue star) calculations
compared with other work (orange squares)
\cite{Marcucci:2011jm,Congleton:1993epm,Congleton:1996epm,Gazit:2008vm,Fujii:1959,Lodder:1965,Delorme:1970,Kim:1970,Donnelly:1973,Mukhopadhyay:1973,Junker:1983}.
The experimental values (dashed gray line) and their error
(shaded region) \cite{Ackerbauer:1998,Deutsch:1968} are included for comparison with the theory predictions.}
\label{fig:rate.comp}
\end{figure}

In addition to the VMC calculation, a GFMC propagation was performed for models Ia and Ia*,
and corresponding results are reported in Table~\ref{tab:gfmc.rates}.
These two models  provided the fastest and slowest VMC partial capture rates for $A=3$ and should 
give an upper and lower limit on GFMC rates. 
The GFMC method removes spurious contamination from the VMC wave functions by
propagating them in imaginary time $\tau$ and should thus provide more reliable results
for these two nuclear Hamiltonians.
Figure~\ref{fig:rate.comp} displays our average VMC results, as well
as both VMC and GFMC results for models Ia and Ia*, compared with experimental data and past theoretical calculations. The GFMC error
is taken to be half the difference between the two available calculations. 

At the VMC level, model Ia overpredicted the $A\,$=$\,3$ muon capture rate by 4.6\%. After propagation, the rate is
decreased and reaches a 1.5\% agreement with the datum. By contrast, model Ia*, which had 1.4\% agreement with the
experimental datum at the VMC level, now underpredicts the rate by 4.2\%.

In Fig.~\ref{fig:rate.comp} panel (a), one sees that the results of
past chiral calculations in Refs.~\cite{Marcucci:2011jm} and~\cite{Gazit:2008vm} fall within the 
bounds for the NV2+3 GFMC rate provided by models Ia and Ia*.  Even when using 
the more recent value of $G_V$, the rate of ~\cite{Marcucci:2011jm} falls within our GFMC band.
While there is this agreement,
because these past calculations use a different set of chiral currents and underlying nuclear interactions than the present work,
it is difficult to directly compare them to our GFMC results. In the future, benchmark calculations with other {\it ab initio}
methods based on the same dynamical inputs would be useful to further validate the present microscopic approach.

While the $A\,$=$\,3$ GFMC rates
exhibit few-percent decreases from the VMC ones,
the $A\,$=$\,6$ rates display a dramatically different behavior for models Ia and Ia*. 
The matrix elements for the model Ia calculation were fairly stable when propagated from VMC to GFMC, resulting in a modest 
sub-percent change of the overall rate.  However, for model Ia*, the dominant matrix elements changed at the few percent level, but since 
the rate is proportional to the square of the matrix element, this leads to a change of roughly $20\%$ in the rate. 

To further understand this behavior, one can look at the system size as a function of $\tau$ during the GFMC propagation of the $A\,$=$\,6$ 
nuclei. The system size for $^6{\rm Li(1^+;0)}$ grows at the same rate in $\tau$ for both models; however, the $^6{\rm He}(0^+;1)$ ground state size is stable for 
model Ia while increasing monotonically in $\tau$ before beginning to converge for model Ia* (see Supplemental Material~\cite{supplement}). 
Because of the ${\rm e}^{-i\bfq\cdot\bfr_{i}}$ dependence in the dominant one-body terms of the current operator, 
the matrix elements at a finite value of $q$ experience a more significant drop for model Ia* due to the diffuseness of $^6{\rm He}(0^+;1)$ with that interaction.
Performing the same analysis for the $A\,$=$\,3$ system, we find that the system size is consistent between both models as a function of $\tau$, explaining the similarity
in their decreasing trend for this partial muon capture rate.

The difference with experiment in $A\,$=$\,6$ is significant for both models Ia and Ia*, especially when compared with
the few percent agreement obtained in GFMC calculations of the GT matrix element for the $^6$He $\to$ $^6$Li beta decay~\cite{King:2020wmp}. 
As detailed in Ref.~\cite{Junker:1983}, calculations of this rate~\cite{Fujii:1959,Lodder:1965,Delorme:1970,Kim:1970,Donnelly:1973,Mukhopadhyay:1973,Junker:1983} 
have ranged from 1160 s$^{-1}$ to 1790 s$^{-1}$.  The calculation of Ref.~\cite{Junker:1983} matched the 
experimental datum by modelling $^6{\rm Li}$ as a $^3{\rm He} + t$ cluster and using the Fujii-Primakoff~\cite{Fujii:1959} effective
Hamiltonian for muon capture. Sub-percent agreement was also obtained by Ref.~\cite{Kim:1970}, which treated the $^6{\rm Li}$ and $^6{\rm He}$ nuclei as elementary particles
with magnetic and axial form factors extracted from experiment. The two calculations presented by the authors of that work adopted different
formulations of the partially 
conserved axial current (PCAC) relation to obtain the pseudoscalar form factor, with the faster rate using the Gell-Mann-L\'{e}vy version~\cite{Gell-Mann:1960}
and the slower rate using the Nambu one~\cite{Nambu:1960}. The Nambu definition is consistent with the induced pseudoscalar term in the weak axial current from 
$\chi$EFT.

It is difficult to compare our result with those of other theoretical
treatments of the $^6{\rm Li}$ partial capture rate, particularly since most of these treatments are decades old. 
For example, in the work of Ref.~\cite{Donnelly:1973} the weak-interaction Hamiltonian is
that of Eq.~(\ref{eq:weak}); however, the $^6$Li and $^6$He bound states are described
by shell-model wave functions with valence configurations
restricted to the $1p$-shell; moreover, the nuclear electroweak
current neglects meson-exchange contributions~\cite{Donnelly:1973}.
We find that our result at leading order (obtained with one-body currents) is
quenched relative to the shell model one, as we would have expected (see Ref.~\cite{King:2020wmp}).
More modern calculations with other {\it ab initio} methods and a novel
measurement of the rate would be valuable in establishing the validity of our
nuclear inputs and many-body approach.

\medskip

\begin{acknowledgments}
This work was supported by the U.S.~Department of Energy, Office of Nuclear Science, under contract DE-SC0021027 (S.~P. and G.~K.), 
a 2021 Early Career Award number DE-SC0022002 (M.~P.),  the U.S.~Department of Energy, Office
of Nuclear Science, under contract DE-AC05-06OR23177 (R.~S.), and by the
FRIB Theory Alliance award DE-SC0013617 (S.~P. and M.~P.) and the U.S.\ Department of Energy NNSA Stewardship 
Science Graduate Fellowship under Cooperative Agreement DE-NA0003960 (G.~K.). The many-body 
calculations were performed on the parallel computers of the Laboratory Computing Resource Center, Argonne National Laboratory, 
and the computers of 
the Argonne Leadership Computing Facility via the 2019/2020 ALCC grant ``Low Energy Neutrino-Nucleus interactions'' for the project NNInteractions,
the 2020/2021 ALCC grant ``Chiral Nuclear Interactions from Nuclei to Nucleonic Matter'' for the project ChiralNuc, and by the 2021/2022 ALCC
grant ``Quantum Monte Carlo Calculations of Nuclei up to $^{16}{\rm O}$ and Neutron Matter" for the project QMCNuc. 
The authors would like to thank J. Carlson, P. Kammel, G. Sargsyan, and L. E. Marcucci for useful discussions. 
\end{acknowledgments}

\bibliography{muon}

\widetext
\textcolor{white}{.}
\newpage

\newpage
\widetext
\begin{center}
\textbf{\large Supplemental Material}
\end{center}

\begin{figure}[tbh]
\begin{center}
\includegraphics[width=0.45\textwidth]{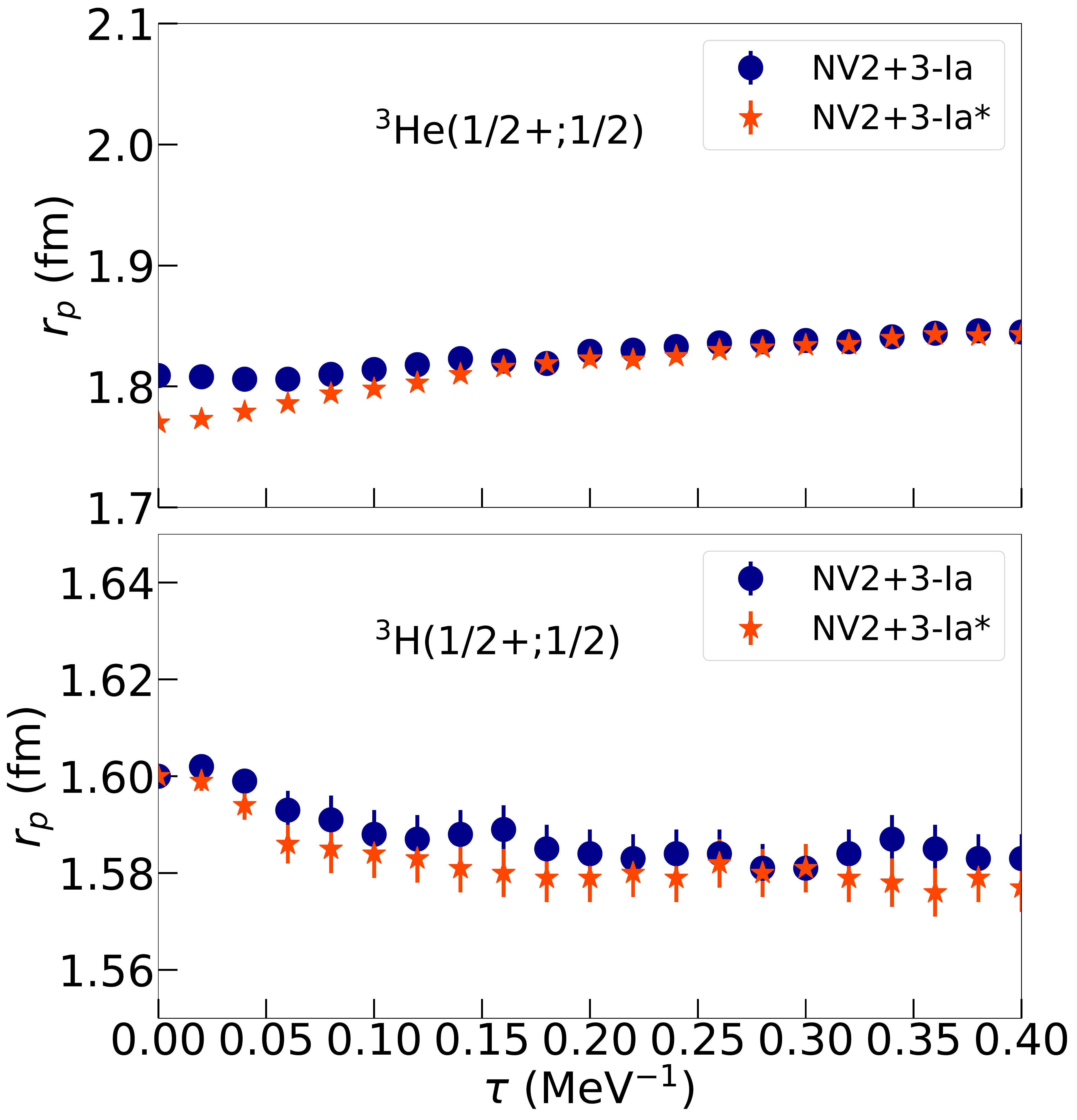}
\end{center}
\caption{The point proton rms radii for the $^3$He and $^3$H  ground states as a function of imaginary time $\tau$ during GFMC propagation. Model Ia (blue circles) and model 
Ia* (orange stars) display similar behaviors for both systems.}
\label{fig:a3.size}
\end{figure}

\begin{figure}[tbh]
\begin{center}
\includegraphics[width=0.45\textwidth]{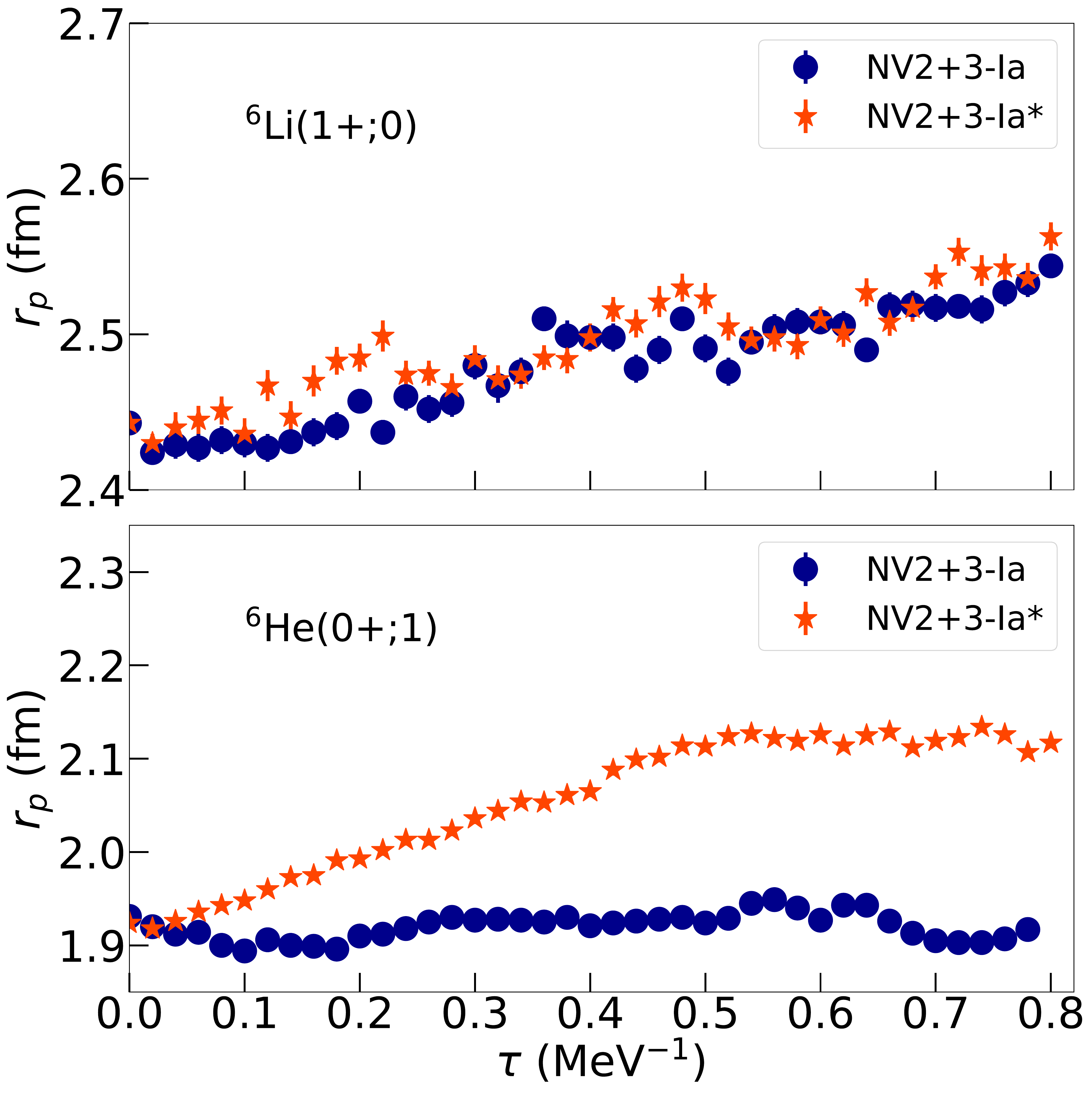}\\
\end{center}
\caption{The point proton rms radii for the $^6$Li and $^6$He ground states as a function of imaginary time $\tau$ during GFMC propagation. Model Ia (blue circles) and model 
Ia* (orange stars) display similar behaviors for $^6$Li, but differ for $^6$He. The model Ia* ground state of $^6$He becomes much more diffuse than its counterpart.}
\label{fig:a6.size}
\end{figure}

This supplemental material is provided to show how the system sizes of the $A\,$=$\,3$ and $A\,$=$\,6$ nuclei evolve during GFMC
propagation in imaginary time $\tau$. As noted in the text, the system size of $^6{\rm He}(0^+;1)$ has a different imaginary time 
behavior for models Ia and Ia*. While it remains stable for model Ia, the system size 
grows monotonically in $\tau$ for model Ia*. Because of this difference and the $e^{i\bfq\cdot\bfr_i}$ dependence of the dominant
one-body current operators, the transition matrix elements for this system experience a larger decrease in model Ia* compared to 
model Ia. This, in turn, leads to the larger drop in the rate going from VMC to GFMC for model Ia* relative to its counterpart. 
For $^6{\rm Li}(1^+;0)$ and the $A\,$=$\,3$ ground states, the system size in $\tau$ is consistent between models Ia and Ia*.
Figures \ref{fig:a3.size} and \ref{fig:a6.size} display the rms point proton radii of the $A\,$=$\,3$ and $A\,$=$\,6$ systems, respectively. 

\begin{figure}[tbh]
\begin{center}
\includegraphics[width=0.45\textwidth]{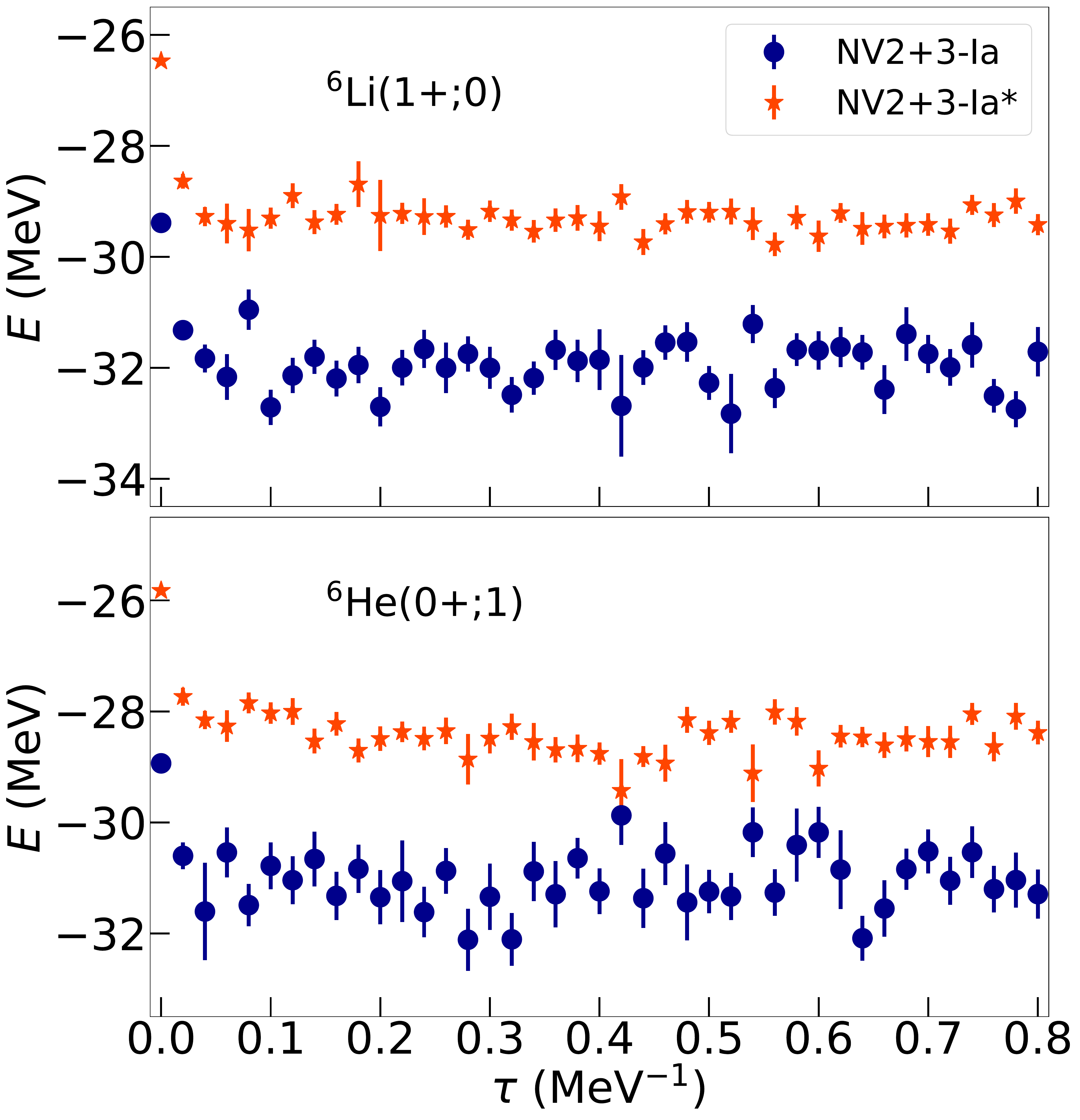}\\
\end{center}
\caption{The energy of the $^6$Li and $^6$He ground states as a function of imaginary time $\tau$ during GFMC propagation. The energies for both model Ia (blue ciricles) 
and model Ia* (orange stars) plateau near $\tau \simeq 0.1$ MeV$^{-1}$.}
\label{fig:a6.eng}
\end{figure}

In addition to the point proton rms radii, we include the GFMC energy propagations for the $A\,$=$\,6$ systems in Fig.~\ref{fig:a6.eng}. 
Note that the energies for both $^6{\rm Li}(1^+;0)$
and $^6{\rm He}(0^+;1)$ plateau near $\tau \simeq 0.1$ MeV$^{-1}$. The difference between the two models is not in the convergence behavior of the energy, but rather in its 
extrapolated value. While the energy values are well converged, the radius of the system is sensitive to small changes in the energy. Model Ia* predicts energies
that are $\sim$1 MeV higher than experimental values (see Table IV of Ref.~[51]). In the case of $^6$He, due to its close proximity to the $\alpha + 2n$ breakup threshold, this leads to clustering and an increased system size. 

\begin{figure}[tbh]
\begin{center}
\includegraphics[width=0.45\textwidth]{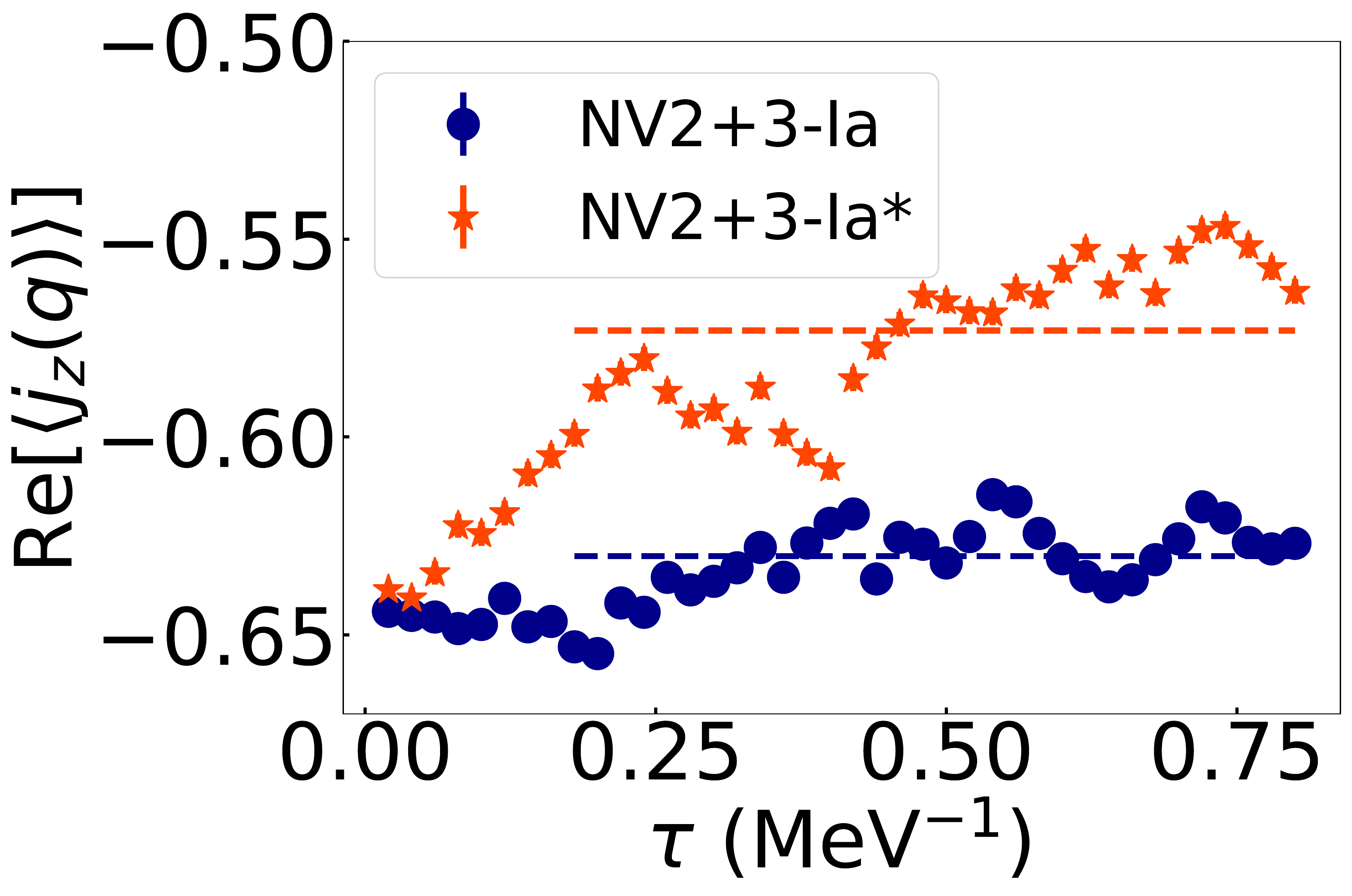}\\
\end{center}
\caption{The GFMC propagation of the real component of the electroweak current operator $j_z$ as a function of imaginary time $\tau$ using model Ia (blue circles) and model 
Ia* (orange stars). The dashed lines indicated the extrapolated value of the matrix element.}
\label{fig:a6.jz}
\end{figure}

We also include an example of an electroweak matrix element propagation in Figure~\ref{fig:a6.jz}. We averaged starting from $\tau = 0.18$ MeV$^{-1}$, after spurious contamination
had been removed from the wave functions and the matrix element was sufficiently converged. While terms like $e^{i\bfq\cdot\bfr_i}$ impact the overall size of the matrix element,
the convergence behavior exhibited by the electroweak matrix element is not the same as that of the radius. Thus, we can safely extract the matrix elements after sufficient propagation in
imaginary time $\tau$. 

\end{document}